\documentclass[final,5p,sort&compress,letterpaper]{elsarticle}

\usepackage{appendix}
\usepackage{graphicx}
\usepackage{caption}
\usepackage{subcaption}
\usepackage{bm}
\usepackage{srcltx}
\usepackage{amsmath}
\usepackage{color}
\usepackage{array}
\usepackage{hyperref}

\allowdisplaybreaks

\newcommand{\be}{\begin{equation}}
\newcommand{\ee}{\end{equation}}
\newcommand{\bea}{\begin{eqnarray}}
\newcommand{\eea}{\end{eqnarray}}

\def\mass{{\tt m}}
\def\charge{{\tt q}}
\def\vpar#1{U_{#1}}
\def\vperp#1{v_{#1_{\perp}}}
\def\LL{\mathcal{L}}

\def\V{\varphi} 
\def\Avec{\mathbf{ A}} 

\def\Np{N_{\rm p}}

\def\PB#1,#2;{\{{#1},{#2}\}}
\def\fd#1,#2;{\frac{\delta{#1}}{\delta{#2}}}

\def\eqref#1{(\ref{#1})}

\journal{COMPHY; doi:10.1016/j.cpc.2014.07.004}

\begin{document}

\begin{frontmatter}

\title{{Application of the phase space action principle to finite-size particle
plasma simulations in the drift-kinetic approximation} }

\author[]{E.~G. Evstatiev\corref{cor1}}
\ead{evstati@mailaps.org}
\address{FAR-TECH, Inc., 10350 Science Center Dr., Suite 150, San Diego, CA 92121}
\cortext[cor1]{Corresponding author}

\begin{abstract} 
We formulate a finite-size particle numerical model of strongly magnetized plasmas 
in the drift-kinetic approximation. We use the phase space action
as an alternative to previous variational formulations based on Low's
Lagrangian or on a Hamiltonian with a non-canonical Poisson bracket.
The useful property of this variational principle is that
it allows independent transformations of particle coordinates and velocities, i.e.,
transformations in particle phase space. 
With such transformations, a finite degree-of-freedom drift-kinetic action is obtained
through time-averaging of the finite degree-of-freedom fully-kinetic action. 
Variation of the drift-kinetic Lagrangian
density leads to a self-consistent, macro-particles and fields numerical model.
Since the computational particles utilize only guiding center coordinates and velocities, 
there is a large computational advantage in the time integration part of the 
algorithm. Numerical comparison between the time-averaged fully-kinetic and drift-kinetic
charge and current, deposited on a computational grid, offers insight into the range of 
validity of the model.
Being based on a variational principle, the algorithm respects the energy conserving
property of the underlying continuous system. 
The development in this paper serves to further emphasize the advantages of using 
variational approaches in plasma particle simulations.

\end{abstract}

\begin{keyword}
Numerical \sep Plasma \sep Kinetic \sep Magnetized  \sep Drift-Kinetic \sep Energy Conserving 
\sep Particle-In-Cell 

\PACS 52.65.-y \sep 52.25.Xz

\end{keyword}

\end{frontmatter}

\section{Introduction}
\label{Introduction}

Finite-size particle algorithms for kinetic plasma simulations
have established a strong record of success in a variety of 
areas\cite{Birdsall:1991aa,Hockney88}.
Recent work~\cite{evstatiev_variational_2013} reexamined the variational formulations
of these algorithms~\cite{lewis:1970:136,Eastwood:1991aa} and developed important improvements 
and generalizations. Two variational formulations were considered 
in Ref.~\cite{evstatiev_variational_2013},
one based on Low's Lagrangian~\cite{low:1958:282}, which generalized previous 
work and a new one, 
based on a Hamiltonian functional and a non-canonical Poisson 
bracket~\cite{morrison:1980:383,morrison:1998:467}.
Ref.~\cite{evstatiev_variational_2013} analyzed in detail the 
relation between Lagrangian symmetries and conservation properties
in the process of reduction from infinite (Vlasov--Poisson or \allowbreak Vlasov--Maxwell system) 
to finite number of degrees of freedom (DOF), 
pointing out which approximations led to the retention -- or not -- of which conserved
quantities. It also addressed the relation between force interpolation and particle shape 
and showed that the particle shape is not the determining factor in an algorithm's 
overall accuracy; as an illustration, it constructed a charge deposition rule that has a 
narrow stencil but high smoothness. Energy conservation properties were 
carefully examined and was shown that in the time-discretized system, energy
conservation depends \emph{only} on the time step size; as a comparison,
it showed that in the more standard 
particle-in-cell (PIC) algorithm energy conservation
depends on both the time 
step and the grid spacing.
There are many other attractive features of variational formulations,  including
the ease of change of variables, 
a consistent way of increasing the overall accuracy (in space and time),
etc., which motivate us to seek their further extensions and applications.

The present work has two purposes. First, to offer an alternative formulation to the above two,
a formulation based on the phase space action~\cite{landau_lifshitz_1,Ye:1992aa}.
In this variational principle the particle coordinates and velocities
are considered independent variables and are varied separately.
The physical relations between coordinates
and velocities as well as Newton's equations of motion are
obtained after performing the variation.
The important feature of this approach is that it allows transformation of variables
independently for coordinates and velocities, i.e., transformations in 
\emph{particle phase \allowbreak space}
rather than in configuration space only. 
This property was used by Littlejohn~\cite{littlejohn_variational_1983},
who offered a simplified derivation of the guiding
center equations of motion of a point particle in
external electric and strong magnetic fields.
Thus, the second purpose of the present work is to
introduce the guiding
center equations of motion for finite size particles in
a self-consistent, particles and fields numerical model. Since
this is a variational formulation, the energy conserving property is
automatically preserved.

We conduct a brief discussion of related literature to
help point out certain novel aspects of our work. 
The early publication of Lee and Okuda~\cite{lee_simulation_1978} 
presented a particle model based on drift-kinetic electrons and fully-kinetic
ions and used it to simulate the linear and non-linear stages of drift-wave instabilities.
An important advantage of such approach was the
large reduction in the computational cost due to the 
larger time step for pushing electrons; another advantage was 
the ability to use realistic electron-to-ion mass ratio. 
In a later publication, further computational efficiency was targeted
by using gyro-kinetic ions in addition to the drift-kinetic 
electrons~\cite{lee_gyrokinetic_1983,lee_gyrokinetic_1987}.
These models were further developed in Refs. \cite{yulin:2005,lin_improved_2011} 
and were used to study magnetic reconnection \cite{wang_particle_2008}.
The review article by Garbet \emph{et al.}~\cite{garbet_gyrokinetic_2010}
provides a summary of other particle-based simulation efforts
and available codes.

The main method followed by the above authors in obtaining a
finite DOF system, i.e., a numerical model,
was to first time- or gyro-phase average the 
fully-kinetic continuous equations to obtain drift- or gyro-kinetic 
continuous equations and \emph{then} apply 
specific spatial and time descretizations. 
This is similar to the approach used in the PIC method.
In doing so, existing conservation laws in 
the continuous system do not automatically transfer to
the resulting numerical model.
For example, the loss of energy conservation is due to errors 
of order higher than the discretization accuracy.
It is known that nonphysical numerical 
artifacts occur in so-discretized 
systems~\cite{okuda:1972:475,Cormier-Michel:2008bs}.
In contrast, our starting point is a finite DOF
fully-kinetic system, i.e., a reduced
Maxwell--Vlasov system described by finite-size particles
and spatially discretized fields (with continuous time).
To this reduced system, we then apply a time-averaging procedure 
to directly obtain a numerical model with finite-size particles and
spatially discretized fields in the drift kinetic
approximation (the gyro-kinetic approximation 
lies outside the scope of the present work).
All steps, including those leading to the reduced fully-kinetic system and its 
time-averaging, are performed within the 
Lagrangian framework, which permits to preserve to the fullest 
the existing symmetries of the original continuous system;
in particular, the energy-conserving property is preserved.
Additional conservation laws may be respected depending 
on the specifics of the discretization~\cite{evstatiev_variational_2013}.
In following this approach, we construct
discretization schemes, in which discretization errors
cancel out exactly to make the conservation of certain quantities possible.
The Lagrangian framework is not mandatory in deriving such discretizations,
however alternatively, one would be faced with the difficult
task of tracking unbalanced errors and modifying 
discretizations to achieve the same effect. 

The energy-conserving deficiency in the PIC model was addressed recently
in two publications~\cite{chen_chacon:2011,markidis_lapenta:2011}.
In addition, a novel implicit technique was introduced that projects
large computational advantage. Only fully-kinetic plasmas were addressed in these works. 

Recent variational finite-size particle formulations were reported by several 
authors~\cite{squire_geometric_2012,xiao_variational_2013,kraus_variational_2013,
shadwick_aps_2013}.
These models were restricted to fully-kinetic electrostatic or electromagnetic plasmas.
Another difference with the present work is that these authors use
time and space discretized action vs. our use of 
continuous time and spatially discretized Lagrangian; in fact,
our keeping time continuous is crucial in order to apply the time-averaging 
procedure to the fully-kinetic Lagrangian (\ref{DK_Averaging}).
The equations of our formulation are most suited
to time-explicit schemes while those in the cited references result, as a rule,
in time-implicit schemes~\cite{Marsden:2001aa} (e.g., when magnetic field is included or 
when higher than second order time integration is desired).

The rest of the paper is organized as follows. Section~\ref{Particle_PS}
describes an alternative formulation of finite-size particle algorithms
based on the phase-space action. Section~\ref{DK} describes
the drift-kinetic approximation of the phase space action and
the drift-kinetic numerical model.
Section~\ref{Tests} provides
numerical comparison between the fully-kinetic and the drift-kinetic
models. Section~\ref{Conclusions} discusses the results
and concludes.


\section{Phase space variational principle}
\label{Particle_PS}


Our starting point is the phase space Lagrangian density (or simply Lagrangian) 
for the fully-kinetic system of particles and fields in Coulomb 
gauge~\cite{landau_lifshitz_1,Ye:1992aa},
reduced to a finite number of degrees of freedom~\cite{evstatiev_variational_2013}:
\begin{align}
\LL_{_{\rm FK}} =&\, \sum_{\alpha=1}^{\Np}\, w_\alpha \left[ {\mass} \mathbf{v}_{\alpha} 
+ {\charge}\Avec(\mathbf x_\alpha,t) \right] \cdot \dot{\mathbf{x}}_\alpha 
\nonumber \\ & 
\!\!- \sum_{\alpha=1}^{\Np}\, w_\alpha \left[ \frac{1}{2}{\mass}\mathbf v_\alpha^2 
+ {\charge} \V(\mathbf x_\alpha,t) \right] \nonumber \\
&\!\! +h_x h_y h_z\left\{-\frac{\epsilon_0}{2}\!\sum_{\bf m,n}\,\V_{\bf m}(t)\nabla_{\bf mn}^2\V_{\bf n}(t)
\right. \nonumber \\ &\left. 
\!\!+\, \frac{\epsilon_0}{2}\!\sum_{\bf m} \dot\Avec_{\mathbf m}(t) \cdot \dot\Avec_{\mathbf m}(t) 
+\! \frac{1}{2\mu_0}\!\sum_{\bf m,n}\! \Avec_{\bf m}(t) \cdot \nabla_{\bf mn}^2\Avec_{\bf n}(t) \right\} \! ,
\label{L_PS}
\end{align}
where $\Np$ is the number of simulation particles; $w_{\alpha}$ is
their computational weight; ${\mass}$ and
${\charge}$ are the physical mass and \allowbreak charge of the plasma species 
(we do not show explicitly a sum over particle species but such can be trivially
added); $\epsilon_0$ and
$\mu_0$ are the permittivity and permeability of vacuum; 
$\V_{\bf m}$ and $\Avec_{\bf m}$ 
denote the collection of grid (or nodal) values
of the electric and magnetic vector potential, respectively, on a three-dimensional
grid with $\mathbf{m}\equiv (m_x,m_y,m_z)$;
the grid is assumed uniform with grid spacings $h_x$, $h_y$, and $h_z$;
sums in $\mathbf{m},\mathbf{n}$ range over all grid points;
$\mathbf{x}_\alpha$ is the computational particle's coordinate
and $\dot{\mathbf{x}}_\alpha$ its time derivative; 
$\mathbf{v}_\alpha$ is the particle's velocity, which
at this point is considered an independent variable, i.e.,
unrelated to $\dot{\mathbf{x}}_\alpha$.
The abbreviated notations $\V(\mathbf x_\alpha,t)$ and $\Avec(\mathbf
x_\alpha,t)$ have been used to denote interpolated values of the
electric and magnetic vector potential from the computational grid to
the particle location: 
\begin{align}
\V(\mathbf x_\alpha,t) &=\, \sum_{\bf m}\,\rho_{\bf m}(\mathbf x_\alpha)\V_{\bf m}(t)\, ,
\label{V_definition} \\
\Avec(\mathbf x_\alpha,t) &=\, \sum_{\bf m}\,\rho_{\bf m}(\mathbf x_\alpha)\Avec_{\bf m}(t)\, .
\label{A_definition}
\end{align}
$\rho_{\bf m}(\mathbf x_\alpha)$ is a charge
deposition rule of choice; $\rho_{\bf m}(\mathbf x_\alpha)$ could either be chosen 
from some of the well known rules in the particle-in-cell method 
\cite{Birdsall:1991aa,Hockney88} or from the more general
ones described in \cite{evstatiev_variational_2013}.
The operator $\nabla^2_{\bf mn}$
denotes the appropriately discretized Laplacian operator, e.g., by central differences.
Additionally, we introduce the following  notation 
[which becomes apparent in deriving
equations \eqref{dot_x}--\eqref{nabla2_A}]:
\begin{align}
{\bf B}(\mathbf x_\alpha,t) &= \nabla\times \Avec(\mathbf x_\alpha,t)  =
\sum_{\bf m}\,\frac{\partial \rho_{\bf m}(\mathbf x_\alpha ) }{\partial \mathbf x_\alpha }
\times \Avec_{\bf m}(t) \, , \label{B_definition} \\
\nabla\V(\mathbf x_\alpha,t ) &  = 
\sum_{\bf m}\,\frac{\partial \rho_{\bf m}(\mathbf x_\alpha ) }{\partial \mathbf x_\alpha } 
\, \V_{\bf m}(t) \, , \qquad \nonumber \\
\frac{\partial A(\mathbf x_\alpha,t)}{\partial t} &= 
\sum_{\bf m}\,\rho_{\bf m}\left( \mathbf x_\alpha\right) \frac{{\rm d}\Avec_{\bf m}(t)} {{\rm d} t}
\, ,  \nonumber \\
{\bf E}(\mathbf x_\alpha,t) &= - \nabla\V(\mathbf x_\alpha,t) 
- \frac{\partial A(\mathbf x_\alpha,t)}{\partial t} \, .  
\label{E_definition}
\end{align}
In the following part of the paper, where we do not show explicitly the arguments of 
field variables, we assume
definitions similar to \eqref{V_definition}--\eqref{E_definition}; we also assume the convention of
summation over repeated indices.

The independent variables in the Lagrangian \eqref{L_PS} are 
$(\mathbf{x}_\alpha,\allowbreak\mathbf{v}_\alpha,\V_{\bf m},\Avec_{\bf m})$.
The equations for the electric and magnetic fields are obtained
by a variation with respect to the corresponding variable, $\V_{\bf m}$ and $\Avec_{\bf m}$. 
Keeping in mind that {independent} variations of particle
positions $\mathbf{x}_\alpha$ and velocities $\mathbf{v}_\alpha$ are performed in \eqref{L_PS},
we obtain
\begin{align}
\dot{\mathbf{ x}}_\alpha &=\, \mathbf {v}_\alpha \label{dot_x}\, , \\
{\mass}\dot{\mathbf{v}}_\alpha &= 
- {\charge}\left[ \nabla\V(\mathbf x_\alpha,t ) 
+ \frac{\partial \mathbf A(\mathbf x_\alpha,t )}{\partial t}\right]
\!+ {\charge}\, \dot{\mathbf{x}}_\alpha\! \times\! \nabla\!\times\! \Avec(\mathbf x_\alpha,t ) \, 
\nonumber \\ 
&=\,\, \charge\, \big[ \mathbf E(\mathbf x_\alpha,t)
+ \dot{\mathbf{x}}_\alpha \times \mathbf B(\mathbf x_\alpha,t) \big]\, ,
\label{dot_v} 
\end{align}
\vspace{-20pt}
\begin{align}
& h_x h_y h_z \nabla_{\bf mn}^2\V_{\bf n} = 
- \frac{\charge}{\epsilon_0} \sum_{\alpha=1}^{\Np}\, w_\alpha \rho_{\bf m}(\mathbf x_\alpha) \, ,
\label{nabla2_V} \\
& h_x h_y h_z\frac{1}{\mu_0}\left[\nabla_{\bf mn}^2\Avec_{\bf n} 
- \frac{1}{c^2}\ddot{\Avec}_{\bf m}\right] =
- {\charge} \sum_{\alpha=1}^{\Np}\, w_\alpha \dot{\mathbf x}_\alpha 
\rho_{\bf m}(\mathbf x_\alpha)\, , 
\label{nabla2_A}
\end{align}
i.e., a self-consistent set of equations for macro-particles and (spatially discretized) fields.
As already stated, the physical relation between coordinates $\mathbf{x}_\alpha$ and velocities
$\mathbf{v}_\alpha$, Eq.~\eqref{dot_x}, is obtained as a result of the variation. 

We note that Eqs.~\eqref{dot_v}--\eqref{nabla2_A} are suitable for the simulation of fully-kinetic 
electromagnetic plasmas, including electromagnetic waves, with the restriction that
the particles have non-relativistic velocities; Ref.~\cite{stamm_ieee_2013} offers
a 1D relativistic formulation with Low's Lagrangian as a starting point.

The energy is given by 
\begin{align}
W_{_{\rm FK}}  =& \frac{1}{2}\sum_{\alpha=1}^{N_{\rm p}} w_\alpha \mass\dot{\mathbf{x}}_\alpha^2 
+\frac{\epsilon_0}{2}h_x h_y h_z \dot{\Avec}_{\bf m}\cdot \dot{\Avec}_{\bf m} 
\nonumber\\ & \mbox{\hspace{-0pt}} 
- h_x h_y h_z \left[ \frac{\epsilon_0}{2}\V_{\bf m}\nabla_{\bf mn}^2\V_{\bf n} +
\frac{1}{2\mu_0}\Avec_{\bf m}\nabla_{\bf mn}^2\Avec_{\bf n} \right].
\end{align}
The proof of the energy-conserving property is straightforward and is herein omitted.


\section{The drift-kinetic approximation}
\label{DK}


The more general transformation of variables in the 
phase space action principle
was exploited by Littlejohn~\cite{littlejohn_variational_1983} to great advantage. 
He presented a mathematically simple and elegant derivation of
the guiding center equations of motion of a point particle in
external electric and strong magnetic fields. The
derivations in Littlejohn's paper may be  repeated with minor
modifications to account for the finite size of computational particles. 
We remind the reader that the
guiding center approximation aims to filter out
the fast gyro-motion and describe 
the particle motion by an ``averaged,'' guiding center motion.
An important assumption is the smallness of the gyro-radius of
a particle compared to the length scale of interest. 
Thus, in deriving the guiding center Lagrangian, certain ordering
between the various 
quantities is assumed. For example, in 
Littlejohn's derivation the $\mathbf E \times \mathbf B$ velocity drift
is assumed to be of the same order as the $\nabla B$ and curvature $B$ drifts.
This imposes a limitation on the magnitude of the electric field.
One may also consider the case of a strong
$\mathbf E \times \mathbf B$ shear 
(see Ref.~\cite{brizard_foundations_2007} and references therein);
then the details of the averaging procedure must reflect such choice.
In the present work we adhere to Littlejohn's choice of ordering.
Last, we assume that the perturbation fields, i.e., fields due to plasma
space charge and plasma currents, are weak so that do not 
violate the drift-kinetic ordering. We do not consider
directions parallel and perpendicular to the magnetic field separately
since we retain the possibility to treat one of the plasma species as fully-kinetic,
i.e., not strongly magnetized.
Ultimately, the validity of a drift-kinetic model lies within the made assumptions.

We obtain (see \ref{DK_Averaging} and Ref.~\cite{littlejohn_variational_1983}) the
Lagrangian of particles and fields corresponding to the lowest, drift-kinetic
ordering in the small parameter as:

\begin{align}
\LL_{\!_{\rm DK}} =&\, \sum_{\alpha=1}^{\Np} w_\alpha \! \left[ {\charge}\Avec 
+ {\mass}\vpar{\alpha} \hat{\mathbf b} \right]
\cdot\dot{\mathbf X}_\alpha + \left(\frac{\mass}{\charge}\right)  
\sum_{\alpha=1}^{\Np}\, w_\alpha \mu_\alpha\dot\Psi_\alpha 
\nonumber \\ & 
- \sum_{\alpha=1}^{\Np} w_\alpha \! \left[ {\charge}\V
+ \mu_\alpha B
+ \frac{\mass\vpar{\alpha}^2 }{2} \right] 
\nonumber \\ & 
+ h_x h_y h_z\left\{ -\frac{\epsilon_0}{2}\V_{\bf m}(t)\nabla_{\bf mn}^2\V_{\bf n}(t) 
\right. \nonumber \\&\left. 
+ \frac{\epsilon_0}{2} \dot\Avec_{\mathbf m}(t)\cdot\dot\Avec_{\mathbf m}(t) 
+\! \frac{1}{2\mu_0}\Avec_{\bf m}(t) \cdot \nabla_{\bf mn}^2\Avec_{\bf n}(t) \right\} \, .
\label{L_PS_DK}
\end{align}
The definitions of the various quantities in \eqref{L_PS_DK} are as
follows: $\mathbf X_\alpha$ is the guiding center coordinate of
particle $\alpha$ and $\dot{\mathbf X}_\alpha$ its time derivative;
$\vpar{\alpha}$ is the particle's velocity parallel to the magnetic field
line (again, in \eqref{L_PS_DK} it is considered unrelated to  $\dot{\mathbf X}_\alpha$); 
$\Psi_\alpha$ is the gyro-phase of the particle;
$B = \sqrt{(\nabla \times\Avec(\mathbf X_\alpha,t) )^2}$ 
is the magnitude of the magnetic field;
$\hat{\mathbf b}(\mathbf X_\alpha,t) =  \mathbf
B(\mathbf X_\alpha,t)/B(\mathbf X_\alpha,t) $ is the unit vector in the
direction of the magnetic field;  $\mu_\alpha =
{\mass}\vperp{\alpha}^2/(2B)$ is the magnetic moment
of the particle with velocity $\vperp{\alpha}$ perpendicular to the magnetic field line
(see also \ref{DK_Averaging}).

The independent variables in the Lagrangian \eqref{L_PS_DK} are\break $(\mathbf
X_\alpha,\vpar{\alpha},\mu_\alpha,\Psi_\alpha,\V_{\bf m},\Avec_{\bf m})$.  One can
easily see that a variation with respect to $\mu_\alpha$ yields the
decoupled equation \break $\dot\Psi_\alpha = ({\charge}/{\mass})B$, 
a variation with respect to the gyro-phase yields the
conservation of the particle's magnetic moment, $\dot\mu_\alpha = 0$,
and a variation with respect to $\vpar{\alpha}$ yields $\vpar{\alpha}
=  \hat{\mathbf b} \cdot \dot{\mathbf X}_\alpha$. 
The remaining variations of the Lagrangian \eqref{L_PS_DK} 
with respect to ${\mathbf X}_\alpha$, $\V_{\bf m}$, and $\Avec_{\bf m}$ yield 
the self-consistent set of equations:
\begin{align}
\dot{\mathbf X}_\alpha =&\,  
\frac{1}{\hat{\mathbf b}\cdot \mathbf B^*} 
\left\{ \vpar{\alpha}\mathbf B^* + \hat{\mathbf b} \times 
\big[(\mu_\alpha/\charge)\nabla B - \mathbf E^* \big] \right\} \, , 
\label{dot_x_DK} \\
{\mass}\dot{\vpar{}}_{\alpha} = & 
- \frac{\charge}{\hat{\mathbf b}\cdot \mathbf B^*} \mathbf B^* \cdot 
\big[(\mu_\alpha/\charge)\nabla B - \mathbf E^* \big] \, , 
\label{dot_U_DK}
\end{align}
\vspace{-20pt}
\begin{align}
& h_x h_y h_z\, \epsilon_0 \nabla_{\bf mn}^2\V_{\bf n} = 
- \charge \sum_{\alpha=1}^{\Np}\, w_\alpha \rho_{\bf m}(\mathbf X_\alpha) \, , 
\label{nabla2_V_DK} \\ &
h_x h_y h_z\,  \frac{1}{\mu_0} \left( \nabla_{\bf mn}^2 \Avec_{\rm n} -(1/c^2)\ddot{\Avec}_{\bf m} \right) =
\nonumber \\ & \mbox{\hspace{20pt}}
- {\charge}  \sum_{\alpha=1}^{\Np}\, w_\alpha \dot{\mathbf X}_\alpha \rho_{\bf m}(\mathbf X_\alpha) 
\nonumber \\  & \mbox{\hspace{20pt}}
+  \sum_{\alpha=1}^{\Np}\, w_\alpha  \nabla\rho_{\bf m} (\mathbf X_\alpha) \times
\frac{{\mass} \left(\dot{\mathbf X}_\alpha\dot{\mathbf X}_\alpha -U_\alpha^2\mathbf I \right)\cdot 
\hat{\mathbf b}(\mathbf X_\alpha) } 
{B(\mathbf X_\alpha)}  \nonumber \\  & \mbox{\hspace{20pt}}
- \sum_{\alpha=1}^{\Np}\, w_\alpha \mu_\alpha \nabla\rho_{\bf m} (\mathbf X_\alpha) \times 
\hat{\mathbf b}(\mathbf X_\alpha)\,  
\label{nabla2_A_DK}
\end{align}
with unity tensor ${\bf I}$ and
\be \mathbf B^* = \mathbf B +{\mass} U_\alpha\nabla\times \hat{\mathbf b}, \quad
\mathbf E^* = \mathbf E - ({\mass/\charge}) U_\alpha
\frac{\partial \hat{\mathbf b} } {\partial t}\, .
\label{B_star_E_star}
\ee
Eqs.~\eqref{dot_x_DK} and \eqref{dot_U_DK} are the well known guiding
center equations of motion. 
Eqs.~\eqref{nabla2_V_DK} and Eq.~\eqref{nabla2_A_DK},
Poisson's and the wave equation, include self-consistently the plasma response.
Notice that these equations contain only the guiding center particle coordinates and
velocities. Therefore, the resulting electric and magnetic fields correspond to the
time-averaged electric and magnetic fields of the full-kinetic model, 
Eqs.~\eqref{dot_x}--\eqref{nabla2_A} (See also Sec.~\ref{Tests}). 
This means that physics phenomena developing on the faster, 
gyro-motion time scale (e.g., for drift-kinetic electrons, the
electron cyclotron resonance wave--particle interaction) are \emph{not} included in this model.

The inclusion of electromagnetic waves, i.e., the wave 
equation \eqref{nabla2_A_DK}, 
is only necessary when the physics
requires it; for example, with drift-kinetic electrons and
fully-kinetic ions, ion cyclotron resonance
heating may be studied. 
Such numerical model may be constructed by adding the particle and interaction parts
of the fully-kinetic Lagrangian~\eqref{L_PS} to the drift-kinetic Lagrangian~\eqref{L_PS_DK}.
When wave--particle resonance physics in not of interest,
the appropriate model is electrostatic and magnetostatic. Such model may be 
obtained by setting $\dot\Avec_{\mathbf m}(t)=0$ in the Lagrangian~\eqref{L_PS_DK}; then
the term $(1/c^2)\ddot{\Avec}_{\bf m}$ in equation~\eqref{nabla2_A_DK} is missing
and each of the three components of the vector potential satisfies a Poisson equation.

Defining $\mathbf M_\alpha ={\mass} \dot{\mathbf X}_\alpha\dot{\mathbf X}_\alpha/B$ and
a projection operator $\mathbf \Pi_\perp = {\bf I}
-\hat{\mathbf b}\hat{\mathbf b}$, the second term on the right-hand-side of \eqref{nabla2_A_DK} can be
written as $\sum_{\alpha}^{N_{\rm p}}w_{\alpha} \nabla\rho_{\bf m} (\mathbf X_\alpha) \times 
\mathbf \Pi_\perp \cdot \mathbf M_\alpha \cdot \hat{\mathbf b} $; thus we
note two magnetic moment contributions to the current, 
Eq.~\eqref{nabla2_A_DK}: one from the intrinsic gyro-motion, $\mu_\alpha$, and 
one from the drifting guiding center motion, $\mathbf M_\alpha$.

The energy of the drift-kinetic model is given by
\begin{align}
W_{_{\rm DK}} =& \sum_{\alpha=1}^{N_{\rm p}} w_\alpha\left[ \frac{\mass}{2}U_\alpha^2 + \mu_\alpha B \right]
+\frac{\epsilon_0}{2}h_x h_y h_z \dot{\Avec}_{\bf m}\cdot \dot{\Avec}_{\bf m} 
\nonumber\\ & 
- h_x h_y h_z \left[ \frac{\epsilon_0}{2}\V_{\bf m}\nabla_{\bf mn}^2\V_{\bf n} +
\frac{1}{2\mu_0}\Avec_{\bf m}\nabla_{\bf mn}^2\Avec_{\bf n} \right].
\label{Energy_DK}
\end{align}
The energy-conserving property is proved in \ref{Energy_conservation_DK}.

\section{Comparison between the full and drift-kinetic models}
\label{Tests}


In this section we study 
the applicability range of the self-consistent drift-kinetic numerical model. 

In view of the extensive previous 
work on the guiding center equations of 
motion~\cite{kruskal_gyration_1958,northrop_adiabatic_1963,littlejohn_guiding_1979},
we do not further test their validity.
Instead, we focus on verifying that the time-averaged charge and current grid densities
in the fully-kinetic equations, i.e., the time-averaged 
right-hand sides of Eqs.~\eqref{nabla2_V} and
\eqref{nabla2_A}, correspond to their drift-kinetic counterparts,
the right-hand sides of Eqs.~\eqref{nabla2_V_DK} and \eqref{nabla2_A_DK}.

Without having to implement a full three-dimensional particle code, here we present
a more limited yet insightful numerical evidence. Namely, we consider the case of
particles gyrating in a uniform magnetic and zero electric fields. 
This means that we exclude velocity drifts from consideration, 
i.e., we set $\dot {\mathbf X}_\alpha =0$. The assumption
of zero parallel velocity does not present a restriction in
a uniform magnetic field since the parallel component of the fully-kinetic velocity
equals the parallel component of the guiding center velocity; 
the parallel current comparison then reduces to comparison of the charge deposition,
as seen from the right-hand side of Eq.~\eqref{nabla2_A} and the first term
on the right-hand side of Eq.~\eqref{nabla2_A_DK}.
Last, we perform only  
single particle comparisons; again, this is not a restriction 
since the charge and 
current are additive in the number of particles
(of course, keeping in mind that errors are also additive).
To summarize, we compare 
the time-averaged fully-kinetic and drift-kinetic charge and current grid depositions
for a single particle in a uniform magnetic and zero electric fields.

\begin{figure}[!t]
\centering
\begin{subfigure}[!h]{0.34\textwidth}
\centering
\includegraphics[width=\textwidth]{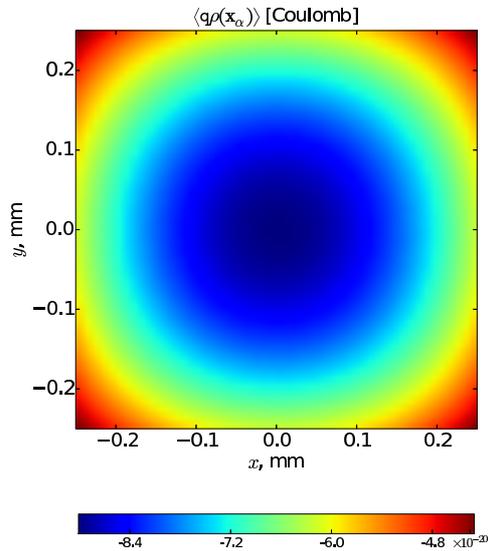}
\caption{Charge deposition by a test electron macro-particle placed at different locations
within the cell. Largest \emph{negative} charge is deposited at the center of the cell.}
\label{fig:Rho}
\end{subfigure}
\begin{subfigure}[!h]{0.34\textwidth}
\centering
\includegraphics[width=\textwidth]{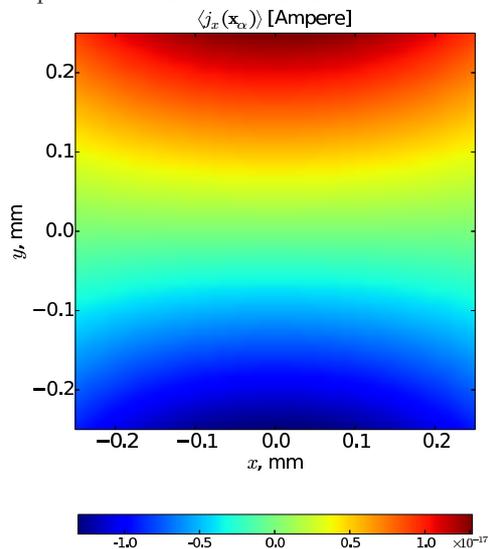} 
\caption{Current deposition $\left<j_x(\mathbf{x}_\alpha)\right>$ 
by a test electron macro-particle placed at different locations within the cell.
Largest values (positive in the upper half) 
of the $x$-component of the current are at the top and bottom
(horizontal) edges of the cell.}
\label{fig:JX}
\end{subfigure}
\caption{
Grid deposition of charge and current. }
\label{fig:Rho_JX}
\end{figure}
\begin{figure}[!t]
\centering
\begin{subfigure}[!t]{0.34\textwidth}
\centering
\includegraphics[width=\textwidth]{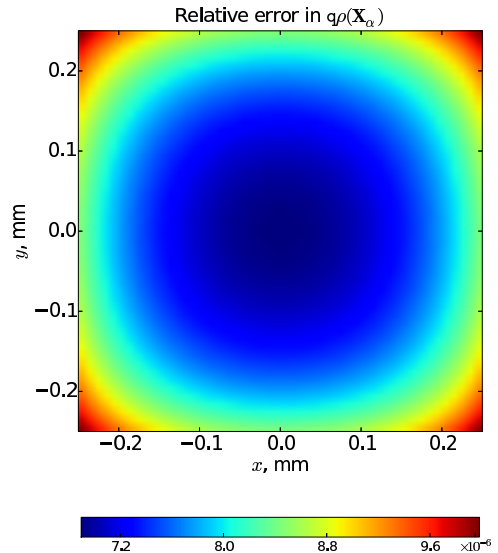}
\caption{Relative error in the charge deposition
by a test electron macro-particle placed at different locations within the cell.
Largest error is at the corners of the cell.}
\label{fig:Error_Rho}
\end{subfigure}
\begin{subfigure}[!t]{0.34\textwidth}
\centering
\includegraphics[width=\textwidth]{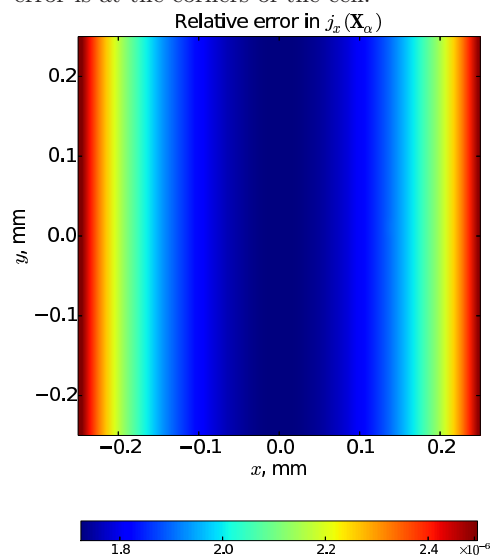} 
\caption{Relative error in the current deposition 
by a test electron macro-particle placed at different locations within the cell. 
Largest error is along the vertical edges of the cell.}
\label{fig:Error_JX}
\end{subfigure}
\caption{ Relative error in the grid deposition of charge and current. }
\label{fig:Error_Rho_JX}
\end{figure}

Let us define the time-average of a quantity $f(\mathbf{x}_\alpha,t)$ as:
\be
\left< f(\mathbf{x}_\alpha) \right> = 
\frac{1}{T}\int_{0}^{T}\!{\rm d}\tau\, f(\mathbf{x}_\alpha,\tau),
\label{def_average}
\ee
where $T$ is the particle's gyro-period. We note that after averaging over
the gyro-period, the quantity $f$ retains time dependence
on the longer, drift-kinetic time scale (time argument not shown in \eqref{def_average}).
We define the relative error of a grid quantity $f$ as
\be
\mbox{Relative error} = \mbox{abs}\left(\frac{\left<f(\mathbf{x}_\alpha)\right> - f(\mathbf{X}_\alpha)}
{\left<f(\mathbf{x}_\alpha)\right>}\right)\, .
\label{def_relative_error}
\ee
This is a measure of how well a drift-kinetic quantity approximates the time-averaged
fully-kinetic quantity \emph{on the computational grid}.
With these definitions, we numerically verify the following equalities:
\nopagebreak
\begin{align}
\left<\charge\,\rho(\mathbf x_\alpha)\right> =&\, \charge\,\rho(\mathbf X_\alpha), 
\label{rho_average} \\
\left<j_x(\mathbf x_\alpha)\right> \equiv&\,
 \left<\charge\,\dot{x}_\alpha\rho(\mathbf x_\alpha) \right>
= \mu_\alpha \frac{\partial \rho(\mathbf X_\alpha) }{\partial Y_\alpha} 
\equiv j_x(\mathbf X_\alpha) \,,
\label{j_x_average}\\
\left<j_y(\mathbf x_\alpha)\right> \equiv&\, 
\left<\charge\,\dot{y}_\alpha\rho(\mathbf x_\alpha) \right>
= -\mu_\alpha \frac{\partial \rho(\mathbf X_\alpha) }{\partial X_\alpha}
\equiv j_y(\mathbf X_\alpha) \,,
\label{j_y_average}
\end{align}
where in the last two equations we have expanded the cross product in the third 
term on the right-hand side of Eq.~\eqref{nabla2_A_DK},
the only non-zero term in our case. We note that 
the current deposition \eqref{j_x_average} and \eqref{j_y_average} 
is \emph{independent} of the sign of the charge;
this is because reversing the sign of the charge reverses the direction of
gyration of the particle, leaving the time-average unchanged
($\mu_\alpha>0$  by definition). 
In the general case  $\dot {\mathbf X}_\alpha\ne 0$ the 
dependence of the drift-kinetic current on the sign of $q$ shows in
the first term on the right-hand-side of \eqref{nabla2_A_DK};
however, notice that the second term in that equation is also 
independent of the sign of $\charge$.

For all numerical tests, we choose $h_x=h_y$ and use a single electron macro-particle
(negative charge and computational weight $w_\alpha=1$).
The magnetic field is oriented in the positive $z$-direction. The charge deposition
rule (interpolating function) is chosen to be a quadratic 
spline; however, in
a full implementation one would desire an interpolating function with \emph{two} continuous
derivatives (second derivative with respect to the particle coordinate $\mathbf{X}_\alpha$ appears
in the term $\nabla B$), such a cubic spline, to ensure continuity of the force as a particle
moves across cell boundaries.

In figure~\ref{fig:Rho_JX} we presents a view of the time averages of the 
fully-kinetic charge and current deposited on the nearest grid point; 
that is, the amount deposited on that grid point as a function of the particle location. 
The particle locations were chosen uniformly within the cell with 
$|x_\alpha|<h_x/2$ and $|y_\alpha|<h_y/2$. The electron macro-particle 
was initialized with the same velocity
at each location, $v_\perp=10^5\,$m/s ($v_x = -v_\perp/\sqrt{2}$, $v_y = v_\perp/\sqrt{2}$), 
in a magnetic field of $0.5\,$T. Although the amount of deposited charge and current
on one grid point changes, the total charge and current, 
which are found by summing the contributions from all grid points, are conserved. 
The time average of the charge deposition is symmetric about the cell center, as expected.
The {time average} of the current, $\left<j_x\right>$ (Fig.~\ref{fig:JX}), is \emph{antisymmetric} about $y=0$ 
since the particle contributes the exact same
current in the positive and negative $y$-directions. Similarly, the current deposition
$\left<j_y\right>$ is antisymmetric about $x=0$ (not shown).

\begin{figure}[!t]
\centering
\begin{subfigure}[!t]{0.34\textwidth}
\centering
\vspace{8pt}
\includegraphics[width=\textwidth]{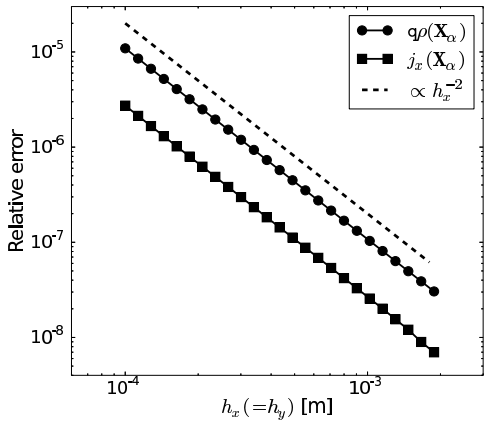}
\caption{Relative error in the charge and current deposition as a function
of grid spacing. }
\label{fig:Error_hxy}
\end{subfigure}
\begin{subfigure}[!t]{0.34\textwidth}
\centering
\vspace{8pt}
\includegraphics[width=\textwidth]{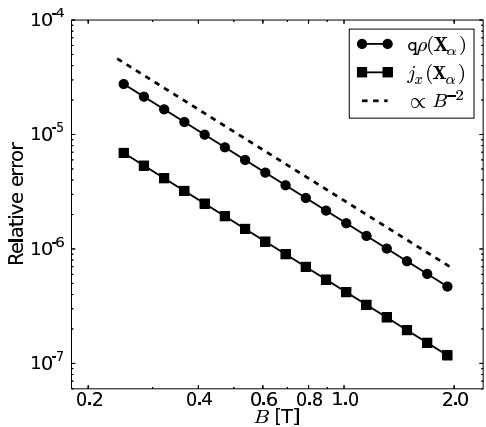}
\caption{Relative error in the charge and current deposition as a function
of magnetic field strength.}
\label{fig:Error_B0}
\end{subfigure}
\caption{ Scaling of relative error. }
\label{fig:Error_hxy_B0}

\end{figure}

Figure~\ref{fig:Error_Rho_JX} shows 
the relative error in the deposited charge  and current 
from figure~~\ref{fig:Rho_JX}, as defined in \eqref{def_relative_error}.
The error in the deposited charge (Fig.~\ref{fig:Error_Rho}) is largest around the corners of the cell and
is minimal at the center. The error in the deposited current (Fig.~\ref{fig:Error_JX})
peaks at the left and right edges of the cell and is symmetric around $x=0$. 
Such variations of the error are related to the particular choice of interpolating function.

Next, figure~\ref{fig:Error_hxy_B0}
shows the variation of the relative errors 
with grid spacing and magnetic field strength. 
We  have fixed the particle location near the center of the cell.
For the dependence on the 
grid spacing shown in figure~\ref{fig:Error_hxy}  
we have set $B_0=2\,$T and $v_\perp= 10^5\,$m/s. For the dependence
on the magnetic field strength, figure~\ref{fig:Error_B0}, we have chosen $h_x=0.5\,$mm.
The relative error in both the charge and current depositions decreases
inversely with increasing grid spacing, $h_x$, and magnetic field strength, $B$. 
This is expected since finite Larmor radius effects become less important for larger $h_x$,
i.e., the gyration of the particle becomes more ``invisible'' to the grid. 
Increasing the magnetic field strength
achieves a similar effect since the gyro-radius decreases inversely with $B$.
Recall that the validity of the drift-kinetic approximation depends on a small parameter,
the ratio of the Larmor radius to the length scale of interest. In our case, the length
scale of interest is given by the grid spacing, $h_x$. Either larger grid spacing 
or smaller gyro-radius decreases the value of this small parameter,
improving the validity of the drift-kinetic approximation; this is reflected in
the decreasing errors in figure~\ref{fig:Error_hxy_B0}.
The particular scaling law of the relative error depends on the choice of 
interpolation rule. In our case, we see an inverse square dependence 
in both Figs.~\ref{fig:Error_hxy} and \ref{fig:Error_B0}.


\section{Discussion and conclusions}
\label{Conclusions}

Using the phase space action principle, we have 
formulated an energy conserving (continuous time), finite-size particle, 
self-consistent numerical simulation model in the drift kinetic
approximation. This work extends previous variational formulations.
The model can be applied to simulate plasmas
in strong magnetic fields where the electron (and possibly the ion)
population of a plasma is highly magnetized and where 
kinetic effects are of importance.
For example, in magnetized plasma discharges 
the electrons have a very non-Maxwellian velocity distribution. 
For such plasmas, fluid descriptions are inappropriate. 
Withing this category falls
the modeling of electron cyclotron resonance ion sources 
(ECRIS) \cite{geller_electron_1996},
where the electron population is strongly non-Maxwellian. 
Partial simulation results with the quasi-3D code SIMPL 
(SIMulation of PLasmas)~\cite{evstatiev_aps_2013}
are in excellent agreement with the tests presented here.
SIMPL utilizes the hybrid model of drift-kinetic electrons and fully-kinetic ions
briefly described in Sec. \ref{DK}.

Errors in the charge and current deposition on the computational grid 
were examined. It was shown that increasing the grid spacing and 
the magnetic field leads to decreasing of the error. This is
an indication that the drift-kinetic approximation becomes
better; it also serves as a guideline for the model applicability.

We note a certain limitation of the drift-kinetic
Lagrangian \eqref{L_PS_DK}. 
It was shown in Ref.~\cite{correa-restrepo_regularization_1985}
that at very large particle velocities and for magnetic fields
with non-vanishing parallel curl, singularities in the
guiding-center velocity and acceleration appear.
The velocities at which this happens are usually very large, possibly larger
than the speed of light for non-relativistic models;
therefore, most applications may be unaffected by this divergence. 
In the cases when this singularity does occur, an appropriate regularization procedure
of the drift-kinetic Lagrangian is available~\cite{correa-restrepo_regularization_1985}
and can be adapted to our numerical model.

The computational advantage
of the method comes from the ability to increase the simulation 
time step by roughly two orders of magnitude; this is because the 
guiding center equations of motion are used, making it unnecessary
to resolve the fast time scale
of gyro-motion. In this model, 
physics developing on the gyro-period time scale is excluded.

We note that continuity of the force on a particle, as it \allowbreak crosses cell boundaries,
is desirable but not mandatory. 
The energy conserving property of models with discontinuous particle 
force is still preserved~\cite{lewis:1970:136,Eastwood:1991aa};
in fact, the energy variation in time is bounded when
symplectic time integrators are used.
The disadvantage is that the error scales  
more poorly with the time step (in the time-discretized model) compared to models using smoother 
interpolating functions~\cite{evstatiev_variational_2013,shadwick_invited_2013}.
Numerical noise 
in simulations with discontinuous force is also higher.
(To draw a parallel with the particle-in-cell method,  
the nearest-grid point charge deposition also has a discontinuous force and 
results in higher numerical noise.)
Ref.~\cite{evstatiev_variational_2013} 
was able to considerably improve the numerical noise and the accuracy of energy conservation
by providing the freedom to use smooth interpolating functions of any order. 
In the present formulation, the cubic spline is the lowest 
order interpolating function that possesses two continuous derivatives 
of $\rho_{\bf m}(\mathbf{X}_\alpha)$.
It is, of course, possible to use interpolating functions of order higher than cubic; however, 
a potential disadvantage is the higher computational load,
which would diminish to an extent the advantage gained by using larger time steps. 
We postpone a more thorough examination
of the trade-offs between the advantages and disadvantages of our model 
for the future, when a full implementation is available.

It is possible to generalize the method to higher order in the small
(drift-kinetic ordering) parameter; however this 
would also require 
higher computational load and may become disadvantageous.
Another possible generalization is constructing hybrid fluid--kinetic 
or fluid--drift-kinetic energy-conserving numerical models for magnetized plasmas;
examples of such formulations for electrostatic plasmas are given
in \allowbreak Refs.\cite{evstatiev_variational_2013,evstatiev_model_2013}.

The presented work underlines the advantages of using variational approaches
to formulations of finite-size particle algorithms.


\section*{Acknowledgments}
This work was supported in part by the U.S. DOE-SBIR program. 
The author thanks B.~A. Shadwick for the many constructive comments.


\appendix

\section{Modification of the Lagrangian averaging procedure for finite size  particles}
\label{DK_Averaging}

In this appendix we outline the differences that arise \break when deriving
the guiding center Lagrangian for finite-size (computational) particles.
In the point particles case, the averaging procedure is effected through changes of
variables and ``gauge'' transformations~\cite{littlejohn_variational_1983}.
The procedure is based on  particular ordering assumptions, small ratio of
gyro-radius to length scale of interest and slow time variation of
the magnetic vector potential and the electric potential. 
The following coordinate and velocity transformations 
are performed in the kinetic Lagrangian~\eqref{L_PS}:
\begin{align}
\mathbf x_\alpha &= \mathbf X_\alpha + 
\vperp{\alpha}\hat{\mathbf a}(\mathbf X_\alpha,t)/B(\mathbf X_\alpha,t)\, , 
\label{transform_x} \\
\mathbf v_\alpha &= U_\alpha \hat{\mathbf b}(\mathbf X_\alpha,t) 
+ \vperp{\alpha} \hat{\mathbf c}(\mathbf X_\alpha,t)\, , 
\label{transform_v}
\end{align}
where the various unit vectors have the relation $\hat{\mathbf a} =
\hat{\mathbf b}\times \hat{\mathbf c}$. In \eqref{transform_x} and \eqref{transform_v}
the guiding center variables, $\mathbf X_\alpha$ and $U_\alpha$, are functions
of the slow time; so is the magnitude of the perpendicular velocity, $\vperp{\alpha}$.
To cancel out fast time scale
terms, one makes transformations of the Lagrangian by adding
the full time derivative of particular, carefully chosen functions; 
such addition does not  change the equations of motion. 

In our case this procedure needs to be performed on the finite-size particles Lagrangian, 
Eq.~\eqref{L_PS}. Since time in \eqref{L_PS} is continuous, i.e., only spatial
discretization has been \break performed, adding a full time derivative of any function
of particle and (discrete) fields does not change the equations of motion.
For example, one of the transformations~\cite{littlejohn_variational_1983} uses the function:
\be
S = -\left[\frac{\hat{\mathbf a}\,  \vperp{\alpha}}{B} \right] \cdot\Avec\, . 
\label{S_point}
\ee
The analogous function in terms of the variables in \eqref{L_PS} becomes:
\be
S = -\left[\frac{\vperp{\alpha}}{B(\mathbf X_\alpha,t)} \right] \hat{\mathbf a}(\mathbf X_\alpha,t)
\cdot\Avec_{\bf m}(t)\rho_{\bf m}(\mathbf X_\alpha)\, .
\label{S_discrete}
\ee
The full time derivative of \eqref{S_discrete} is evaluated as:
\begin{align}
\frac{{\rm d} S}{{\rm d} t} =&\, 
-\left[\frac{\hat{\mathbf a}\,\vperp{\alpha}}{B}\right] \cdot \Avec_{\bf m}\rho_{\bf m}(\mathbf X_\alpha) \, 
\left[ \dot{\mathbf { X}}_\alpha \cdot \nabla\rho_{\bf m}(\mathbf X_\alpha) \right] 
\nonumber \\ & 
-\left[\frac{\hat{\mathbf a}\, \vperp{\alpha}}{B}\right] \cdot \dot\Avec_{\bf m}(t)\rho_{\bf m}(\mathbf X_\alpha) 
\nonumber \\
& - \frac{{\rm d} }{{\rm d} t} \left[ \frac{\hat{\mathbf a}\,\vperp{\alpha}}{B} \right]
\cdot \Avec_{\bf m}(t) \, \rho_{\bf m}(\mathbf X_\alpha) 
\, . 
\label{dt_S}
\end{align}
Expression \eqref{dt_S} in conjunction with the assumed ordering of 
time scales is then used to obtain the averaged
drift kinetic Lagrangian \eqref{L_PS_DK};
we refer the reader to Ref.~\cite{littlejohn_variational_1983}
for details.

\section{Proof of the energy conserving property of the drift-kinetic model}
\label{Energy_conservation_DK}

In this appendix we prove explicitly that the drift-kinetic approximation model, 
Eqs.~\eqref{dot_x_DK}--\eqref{nabla2_A_DK}, conserves energy when time is kept continuous.
In a time-discretized model energy will not be strictly conserved;
however, symplectic time integrators make its variation bounded and 
decreasing with smaller time step.

The following is a useful auxiliary identity, obtained by crossing Eq.~\eqref{dot_x_DK}
with $\mathbf{B}^*$, rearranging, and dotting with $\dot{\mathbf{X}}_\alpha$:
\begin{align}
\frac{U_\alpha}{\mathbf{\hat b}\cdot\mathbf{B}^*} 
\mathbf{B}^*\cdot\left[\frac{\mu_\alpha}{\charge}\nabla B - \mathbf{E}^*  \right] =& \,
\frac{\mu_\alpha}{\charge}\dot{\mathbf{X}}_\alpha\cdot\nabla B 
- \dot{\mathbf{X}}_\alpha\cdot\mathbf{E}^*.
\label{Aux_U_dot}
\end{align}
The next two useful identities follow from definition \eqref{B_definition}:
\begin{align}
\dot{\Avec}_{\bf m}\cdot\nabla\rho_{\bf m}\times\dot{\mathbf{X}}_\alpha 
=& - \dot{\mathbf{X}}_\alpha\cdot\frac{\partial \mathbf{B}}{\partial t} \, ,
\label{Aux_B_dot_X}\\
\dot{\Avec}_{\bf m}\cdot\nabla\rho_{\bf m}\times{\hat{\mathbf{b}}}
=& - \hat{\mathbf{b}}\cdot \frac{\partial \mathbf{B}}{\partial t} 
\equiv - \frac{\partial {B}}{\partial t}\, .
\label{Aux_dot_B}
\end{align}

The time derivative of the energy \eqref{Energy_DK} is
\begin{align}
\frac{{\rm d}W_{_{\rm DK}}}{{\rm d}t} = &\, \sum_\alpha w_\alpha \left[\mass U_\alpha\dot{U}_\alpha 
+ \mu_\alpha \frac{{\rm d}B}{{\rm d}t}  \right]
\nonumber \\ &
- h_x h_y h_z \epsilon_0\V_{\bf m}\nabla_{\bf mn}^2\dot{\V}_{\bf n} 
\nonumber\\ & 
+ h_x h_y h_z\frac{1}{\mu_0}\dot{\Avec}_{\bf m}\cdot\left[\frac{1}{c^2}\ddot{\Avec}_{\bf m}
-\nabla_{\bf mn}^2\Avec_{\bf n} \right] \, .
\label{dot_W_1}
\end{align}
In Eq.~\eqref{dot_W_1}, we use Eq.~\eqref{dot_U_DK} with Eq.~\eqref{Aux_U_dot}, 
expand the full time derivative of $B$
and use the definition of $\mathbf{E}^*$ from Eq.~\eqref{B_star_E_star}, 
use Eq.~\eqref{nabla2_A_DK} and the time derivative of Eq.~\eqref{nabla2_V_DK} to obtain
\begin{align}
\frac{{\rm d}W_{_{\rm DK}}}{{\rm d}t} =&  \sum_\alpha w_\alpha \left[ -\mu_\alpha\mathbf{X}_\alpha\cdot\nabla B 
+\charge\mathbf{X}_\alpha\cdot\left(\mathbf{E}
-\frac{\mass}{\charge}U_\alpha\frac{\partial\hat{\mathbf{b}}}{\partial t}\right)
\right. \nonumber \\ & \mbox{\hspace{48pt}} \left. 
+ \mu_\alpha\frac{\partial B}{\partial t}+\mu_\alpha\mathbf{X}_\alpha\cdot\nabla B\right]
\nonumber \\ & \mbox{\hspace{0pt}}
+ \sum_\alpha w_\alpha\charge \dot{\mathbf{X}}_\alpha\cdot\left[\V_{\bf m}\frac{\partial{\rho_{\bf m}}}{\partial\mathbf{X}_\alpha} + \rho_{\bf m}\dot{\Avec}_{\bf m} \right]
\nonumber \\ & \mbox{\hspace{0pt}}
+ \sum_\alpha w_\alpha\frac{\mass U_\alpha}{B} \left(\dot{\mathbf{X}}_\alpha\cdot\frac{\partial\mathbf{B}}{\partial t}
- U_\alpha\frac{\partial B}{\partial t} \right)
\nonumber \\ & \mbox{\hspace{0pt}}
- \sum_\alpha w_\alpha\mu_\alpha\frac{\partial B}{\partial t} 
\nonumber  \\   = &  
\sum_\alpha w_\alpha \mass U_\alpha \!
 \left[ 
\frac{1}{B} \left(\dot{\mathbf{X}}_\alpha\cdot\frac{\partial\mathbf{B}}{\partial t}
- U_\alpha\frac{\partial B}{\partial t} \right) 
- \dot{\mathbf{X}}_\alpha\cdot \frac{\partial\hat{\mathbf{b}}}{\partial t}
\right]
 \nonumber \\ 
 =& \, \,  0\, 
\end{align}
after a number of obvious cancellations, using definition \eqref{E_definition}, and the last equality following 
from expanding
\begin{align}
\frac{\partial\hat{\mathbf{b}}}{\partial t} 
\equiv \frac{\partial}{\partial t}\left( \frac{\mathbf{B} }{B} \right) 
= \frac{1}{B}\left(\frac{\partial \mathbf{B}}{\partial t} - \frac{\partial B}{\partial t}\hat{\mathbf{b}} \right)\,.
\nonumber
\end{align}


\end{document}